\documentclass{llncs}

\usepackage{graphicx}	

\begin{document}

\title{Virtual Environments for Training: from individual learning to collaboration with humanoids}
%
\author{St\'ephanie Gerbaud\inst{1} and Nicolas Mollet\inst{2} and Bruno Arnaldi\inst{3}}

\institute{IRISA/INRIA/INSA Rennes, France. stephanie.gerbaud@irisa.fr
        \and
        IRISA/INRIA Rennes, France. nicolas.mollet@irisa.fr
        \and
        IRISA-INSA Rennes, France. bruno.arnaldi@irisa.fr}



\maketitle

\begin{abstract}

The next generation of virtual environments for training is oriented towards collaborative aspects. Therefore, we have decided to enhance our platform for virtual training environments, adding collaboration opportunities and integrating humanoids. In this paper we
put forward a model of humanoid that suits both virtual humans and representations of real users, according to collaborative training activities. We suggest adaptations to the scenario model of our platform making it possible to write collaborative procedures. We introduce a mechanism of action selection made up of a global repartition and an individual choice. 
These models are currently being integrated and validated in GVT\footnote{GVT is a trademark of NEXTER-Group}, a virtual training tool for maintenance of military equipments, developed in collaboration with the French company NEXTER-Group.\newline


\keywordname{ collaborative virtual environments, virtual reality, training, collaboration, virtual humans}
\end{abstract}

\section{Introduction - Context}
The need for virtual environments for training is not to be demonstrated anymore: lower costs and risks, no need for available equipments to train on and control of pedagogical situations are examples of the numerous benefits brought by the use of Virtual Environments (VEs). Among the existing VEs for training, only a few provide collaborative training opportunities. 
However, VEs can bring other assets for collaborative training: 
the possibility to collaborate with distant people via the Internet and the opportunity to train with virtual humans, so that the trainee can train not only everywhere but also at any time.
Therefore, we have decided to enhance our platform for virtual training environments, by adding collaboration opportunities and integrating humanoids. This platform has been specifically designed for training on procedures (especially industrial maintenance procedures) rather than on technical gestures. We are now interested in collaborative procedures.

\begin{figure}
   \begin{minipage}[c]{.46\linewidth}
      \includegraphics[width=6cm]{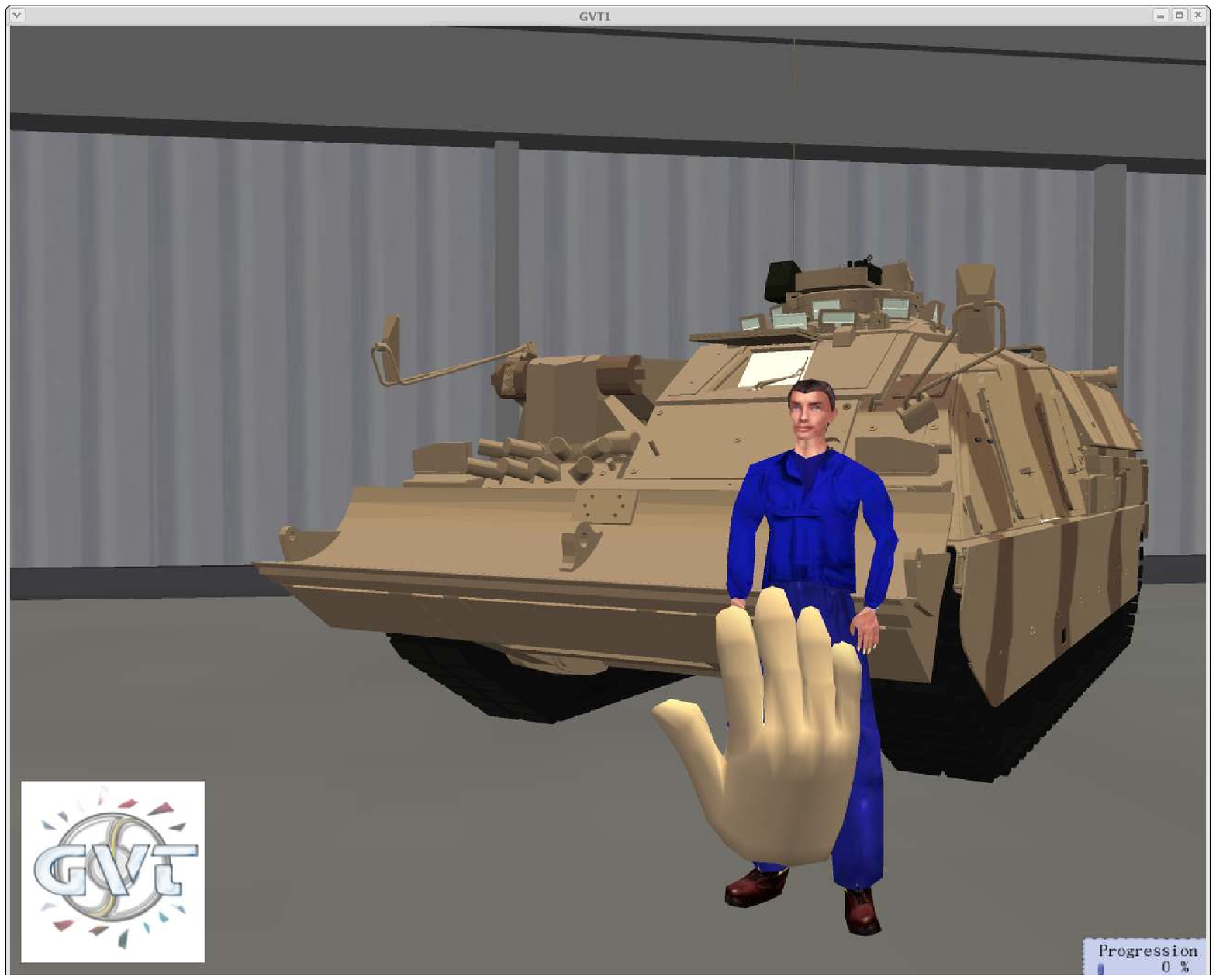}
      \caption{a humanoid in GVT}
      \label{fig:humano}
   \end{minipage} \hfill
   \begin{minipage}[c]{.46\linewidth}
      \includegraphics[width=6.4cm]{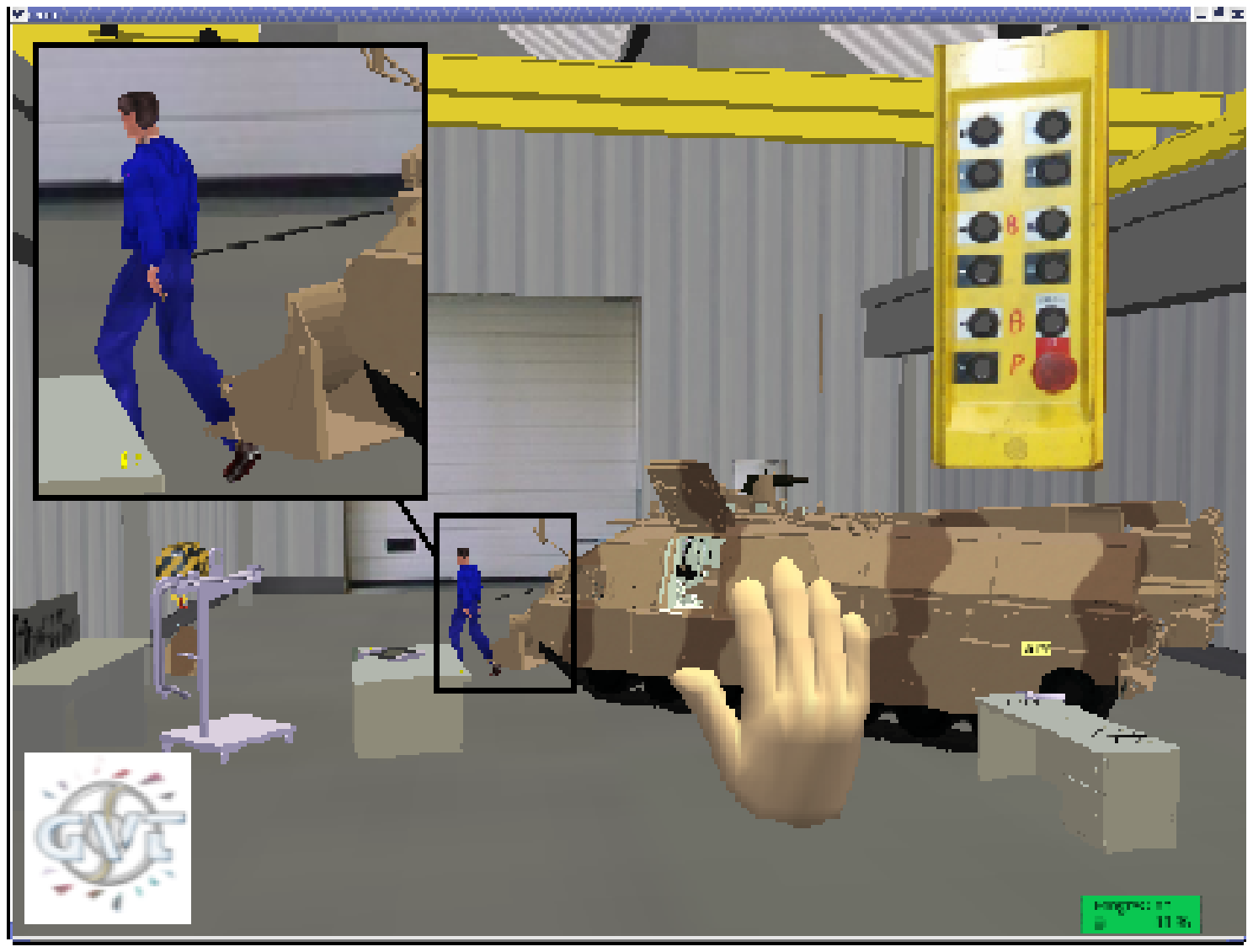}
      \caption{a humanoid interacting in GVT}
      \label{fig:humano2}
   \end{minipage}
\end{figure}

The work presented in this paper is both the integration and the continuation of previous research activities performed in our team (BUNRAKU, at IRISA) dedicated to humanoid animation 
\cite{Menardais2004}, humanoid behavior \cite{Lamarche2002} and collaboration 
\cite{Duval2006}.
In OpenMASK\footnote{http://www.openmask.org}, our Virtual Reality platform, existing features are already deployed to enable collaboration, but at the interaction level. We rely on these facilities to model a higher level of collaboration: collaboration of activity that occurs at the scenario level.
After analyzing the existing Collaborative Virtual Environments (CVEs) for training and the functionalities they must offer, we will first expose the models that we put forward and then finish presenting GVT\footnotemark[4], the industrial application in which they are being integrated (\figurename \ref{fig:humano} and \ref{fig:humano2}).


\section{Related work: CVEs for training}
The next generation of VEs for training is oriented towards collaborative aspects. Collaboration in a VE for training implies sharing: sharing of a common environment, sharing of the others' vision, sharing of activity (the procedure). Those concepts will be reviewed in the following state-of-the-art.
We will first present existing CVEs for training, then we will expose the awareness issue, followed by the requirements for collaborative scenarios and last of all we will describe the use of virtual humans in CVEs for training.

\subsection{Overview of CVEs for training}
Among existing CVEs for training, we will only describe the most relevant ones.\\
COVET (COllaborative Virtual Environment for Training) \cite{Oliveira2000,Oliveira2000a,Hosseini2001} is a prototype of multi-user teletraining application which allows users, represented by avatars, to learn how to replace a faulty card on an ATM switch. It remains a very basic CVE for training since only one user can interact with the objects in the scene (for example the trainer who demonstrates the procedure). The others can only watch this user's avatar acting, change their points of view, move and chat.\\
Steve\cite{Johnson1997,Rickel1999a} has been extended to support team training \cite{Rickel1999,Rickel2003}: students must learn their individual roles in the team and how to coordinate their actions with their teammates. 
Steve agents can play two pedagogical roles: either tutor or substitute for missing teammates.\\
The MRE (Mission Rehearsal Exercise)\cite{Traum2003,Swartout2005} system is a VE for training which has been designed to teach critical decision-making skills to small-unit leaders in the U.S. Army. 
In this application, there is only one trainee but he must collaborate with virtual humans able to converse, show emotions and reason.\\
S\textsc{ecu}R\textsc{e}V\textsc{i} (Security and Virtual Reality) \cite{Querrec2003} is a CVE to train firemen officers for operational management and commandment. This application is based on the MASCARET \cite{Buche2003} model that organizes the interactions between agents and provide them reactive, cognitive and social abilities. 
The goal is to train teams to collaborative and procedural work.

\subsection{Awareness}
The first important need in a CVE is awareness: a user must be aware of the other users populating the environment and their actions. A common solution is to embody the user into an avatar, which can be humanoid. Benford et al. \cite{Benford1995} identified issues that the use of an avatar can solve in a CVE: presence, location and identity being the most fundamental ones. Fraser et al. \cite{Fraser1999,Fraser2000} introduced means of representing the user activity in a CVE. For example, they proposed to use a wire-frame or transparent pyramid of vision to visualize the field of view of one user. They also proposed to extend the avatar's arms to touch the artefact (wire-framed) when portraying the grasping/moving of that object.
As we have seen, several possibilities exist to deal with awareness in a CVE but the avatar appears to be the most meaningful one: "without sufficient embodiment, users only become known to one another through their (disembodied) actions; one might draw an analogy between such users and poltergeists, only visible through paranormal activity" \cite{Benford1995}.
Several types of avatars can be used to achieve these goals, from the most symbolic one, such as a head and two hands to symbolize a real trainee in Steve or a simplified body in COVET, to the most realistic one, such as the humanoids in S\textsc{ecu}R\textsc{e}V\textsc{i} and MRE.

\subsection{Scenario}
Existing ways of writing procedures can be found in \cite{Mollet2006}, but for collaborative procedures an essential notion remains missing: the notion of role. Indeed, we must specify who can do each action and this assignment is generally achieved by associating one role to each action of the scenario, like in MASCARET or Steve. The concept of role (for example the lieutenant role, played by the trainee, in MRE) accounts for the agent's responsibilities in the team.
The role that a student plays determines the actions that this student can or must perform in the procedure. 
Nevertheless, a limitation can be noticed in these environments: only a single role can be associated with a given action of the scenario. Moreover, even though it is not forbidden that more than one person play the same role, the repartition of the actions between several persons playing the same role is not specified and will thus be uncertain: it could always be the same person who performs all the actions, the first person that has the will to do it, etc. In short, there will be no logic in the task repartition. As a result, we can say that these environments are designed for the case where one action of the scenario can be done only by one person, so that each person can always know whether it's up to him to perform this action or not.

\subsection{Virtual humans}
Finally, a useful functionality that is often offered in CVEs for training is the presence of virtual humans that can play specific roles (in MRE) or replace a missing teammate (in Steve or S\textsc{ecu}R\textsc{e}V\textsc{i}).
Virtual humans enable the trainee to realize a collaborative procedure even if nobody else is available. In most of the VEs offering this facility, the virtual human who is a substitute for a missing teammate respects his role and performs the actions of the procedure he has to (Steve, S\textsc{ecu}R\textsc{e}V\textsc{i}). 
In addition, independently of his role in the team, a virtual human could play various pedagogical roles, such as those proposed by Buche et al. \cite{Buche2003}: tutor, companion or troublemaker.

\subsection{Synthesis}
As we have observed in this state-of-the-art, three major functionalities can be identified in a CVE for training: the perception of the others and their actions, the writing of collaborative scenarios with identified roles and the presence of virtual humans. 
However, some limitations can be seen in the existing CVEs for training. These environments only allow to specify one role for each action in the scenario, so that the repartition of actions between the team members is fixed.
Furthermore, virtual humans are predictable because they always follow the procedure, without making errors.

\section{Contribution: models to support humanoids and collaboration}
To palliate the limitations of existing CVEs while preserving their requirements, we propose scenarios that have a strict scheduling of key actions but a flexible repartition of these actions between the team members. Moreover, our virtual humans have a pedagogical profile and, depending on it, may choose to do the correct action in the scenario or another action. Our proposed models are currently being developed under our industrial project, GVT, which is detailed in the next section.

We will begin with an overview of the actual kernel of our platform for virtual training environments, then we will present the global humanoid module, afterwards we will focus on adaptations of the scenario model and we will finish with detailing the mechanism of action selection.

\subsection{The actual kernel of the platform} 
\begin{figure}
\centering
\includegraphics[width=10cm]{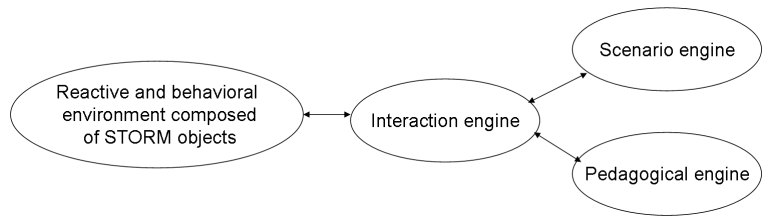}
\caption{Global vision of our platform} 
\label{fig:globalGVT}
\end{figure}

The actual kernel of our platform is divided into four elements, which can be seen on \figurename \ref{fig:globalGVT}:
\begin{itemize}
\item \textbf{a reactive and behavioral environment.} It is an informed 3D world, made of behavioral objects which can interact with each other. The model of behavioral objects and interaction used is STORM (Simulation and Training Object-Relation Model) \cite{Mollet2005}. 
\item \textbf{an interaction engine.} This engine, the STORM engine, aims at managing complex interactions between STORM objects. These interactions depend on the abilities of the STORM objects. This engine is used to determine what could be done in the environment, what are the possible interactions.
\item \textbf{a scenario engine.} This engine is used to know what are the next steps of the procedure that the trainee can do, and its state evolves as the trainee does achieve actions. The scenario is written in the LORA language (Language for Object-Relation Application), whose graphical version is inspired from Graphcet, and which makes the writing of complex procedures possible. More information on LORA can be found in \cite{Mollet2006}.
\item \textbf{a pedagogical engine.} This engine, employed to assist the trainer, uses the two engines above to decide what the trainee is allowed to do. It adapts its reactions to the trainee, and to the pedagogical strategy used. This engine is developed in the CERV\footnote{http://www.cerv.fr/} laboratory. 
\end{itemize}
These elements have been designed to be generic and reusable. In order to build a VE for training, we must: design STORM objects (or re-use existing ones) and specify their abilities, describe the procedure in LORA and define a pedagogical strategy.

\subsection{Modelling of humanoids}
A humanoid, used to reinforce users' awareness, is either the representation of a real user in the virtual world (an avatar) or a virtual human (controlled by the system).
Moreover, the modelling of a humanoid can be divided into two parts: a bodily part and a behavioral part. 
The physical part consists in the visualization and the animation of the humanoid avatar. We use MKM\footnote{http://www.irisa.fr/bunraku/MKM/} (Manageable Kinematic Motion) \cite{Menardais2004} for the appearance of the humanoid and its movements. MKM is a real-time animation engine for synthetic humans which automatically synchronizes, blends and adapts captured motions.
This physical part is encapsulated into the behavioral part, which drives it. A humanoid is a STORM object so that it is possible to interact with it in the same way as with any other object, depending on this humanoid's abilities. For example, on \figurename \ref{fig:humano2} we can see a humanoid interacting with a cable. A humanoid has two kinds of abilities: general abilities owned by every humanoid and specific additional abilities due to the roles he plays.
The behavioral aspect of a humanoid driven by a real human is limited to his roles and his abilities. It is up to the real user to decide what to do. A virtual human has, additionally, a pedagogical profile and a mechanism of decision-making described in section \ref{decision-making}.

\subsection{Adaptation of the scenario model}
In order to make possible the writing of a collaborative scenario, we must adapt the LORA language \cite{Mollet2006}, making modifications that concern:

\subsubsection{Roles allowed for an action}
We must specify what are the roles allowed to perform an action. Therefore we associate the roles allowed and their priorities with each action of the procedure. 
Contrary to other existing CVEs, we allow several roles for each action, which makes some interesting properties appear:
	\begin{itemize}
	\item the task repartition is not fixed anymore: the scenarios are more flexible.
	\item we can use GVT in order to design or perfect procedures. Indeed, thanks to the mechanism of action selection that will be presented in section \ref{action selection}, it is possible to let virtual humans realize the procedure.
	The observation of virtual humans adapting to unexpected situations, the problems they have to cope with or simply the division of work could help to perfect a procedure (change the repartition of tasks, the number of people needed, etc ...). 
	\item implicit collaboration can emerge, such as a virtual human who could help a real user even if it is not explicitly written in the scenario. Indeed, we can specify that an action can be done by different roles or even by anyone that have the suitable abilities. 
	As a result, different persons can be allowed to perform some actions, even if a particular role is expected in priority.
	Thus, if the user with the priority role for an action finds himself in a situation where it is impossible or inconvenient to achieve it, another user may help him and perform this action. 
	Let us take an example: a screw that maintains an object must be unscrewed. The procedure is to hold the object and with the other hand to unscrew.
	This sequence of actions must be done in priority by worker1 or secondary by anyone. 
	We want to realize this scenario in the dark. Therefore, the trainee that plays the role of worker1 has to hold a light all along the scenario, so that he has only one available hand.
	He begins to unscrew the screw and a virtual human that is nearby come to hold the object maintained by the screw because this virtual human has noticed that the trainee wasn't able to perform this action. This is what we call implicit collaboration which is made possible by the mechanism of action selection.
	\end{itemize}

\subsubsection{New types of actions: actions of communication and collaborative actions} 
\begin{figure}
\begin{center}
\includegraphics[width=11cm]{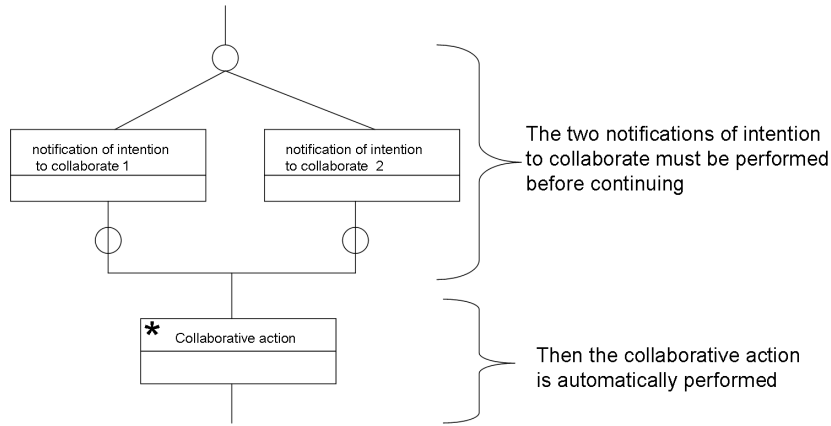}
\caption{representation of a collaborative action in the LORA language} 
\label{fig:coll}
\end{center}
\end{figure}

We must add new types of actions such as actions of communication, to model a communication act that has an important impact on the procedure, and collaborative actions, to model actions made by several users at the same time:
	\begin{itemize}
	\item The communication is sometimes so important in a procedure that this action is an integral part of the procedure (to give an order, to report, etc). For example, when a guide must help his partner to manoeuvre, the guide must communicate information to his partner, in a codified manner. The communication act is transcripted in the scenario by an action of communication.
	This action can be done by selecting the user to communicate with and the message to transmit (for example: "turn on the right").
	\item A collaborative action is an action that must be done by several users at the same time. Such an action, for example holding a heavy object, requires a high synchronization level between the collaborators. The first step is, for each user, to declare his intention to collaborate. In the real life this declaration is done by saying that we are waiting for someone to help us or by beginning the action to show our need for help. 
	This mechanism will be transcripted in the scenario thanks to a special action: \textit{notification of intention to collaborate}. This action, one for each collaborator, will serve for a humanoid to declare that he is ready to collaborate.
	Once every collaborator have done this declaration, the collaborative action automatically begins as we can see on \figurename \ref{fig:coll}.
	Nevertheless, we must set up a timeout in order to cancel the effective notification of intention to collaborate if another notification of intention to collaborate is still missing after this time-limit.
	\end{itemize} 

\subsubsection{Action of grasping and humanoid's hands}
The maintenance activity is based on actions made thanks to tools that the worker holds in his hands. The management of the state of the worker's hands and the actions of grasping and laying of tools are thus omnipresent in maintenance procedures.

\begin{itemize}
\item 
In collaborative scenarios and especially adaptive scenarios where we don't know who will realize some actions, it is very constraining to specify every actions in the scenario like taking and putting a tool. 
In order to make the scenarios more flexible and easier to write, 
the actions of catch and laying of tools are not written in the scenario anymore.
A resource manager deduces the catch of a tool from an action in the scenario that requires this tool, and proposes the laying of an object in order to free the humanoid's hands.
Furthermore, in reality, such actions are implicit in the specification of a maintenance procedure. Thus, making it implicit also in the virtual scenario would make it fit better to the real written procedure.

\item
Information that we must make explicit for each action is the state of humanoid's hands required and their states after the action. Four states are available: free, holding an object, busy and indifferent. 
A resource manager reasons on these states in order to predict if a humanoid is entering a sequence of actions that he won't be able to finish alone (for example because it will monopolize his two hands), so that an implicit collaboration is required.
\end{itemize}

\subsection{Mechanism of action selection}
\label{action selection}
When a virtual human takes part in a collaborative procedure, he must regularly choose the next action he will do, taking into account a lot of parameters.
To manage the behavior of the entities that populate a VE, two approaches exist: a centralized one (example:\cite{Ridsdale1990}), where a global mechanism manages the behavior of each entity and gives them orders, and a local one (example: oRis \cite{Rodin1999}), where each entity is responsible for its own behavior and the global behavior of the environment is emergent (for example multi-agent system and autonomous virtual humans).
In our applicative domain (industrial maintenance training), we must deal with a procedure to respect and virtual humans with a part of autonomy. That is why
we have chosen to mix these two approaches, by modelling autonomous virtual humans that are incited to follow the scenario. Furthermore, this approach tends to reduce the combinatorics of possible actions.

\begin{figure}
\begin{center}
\includegraphics[width=8cm]{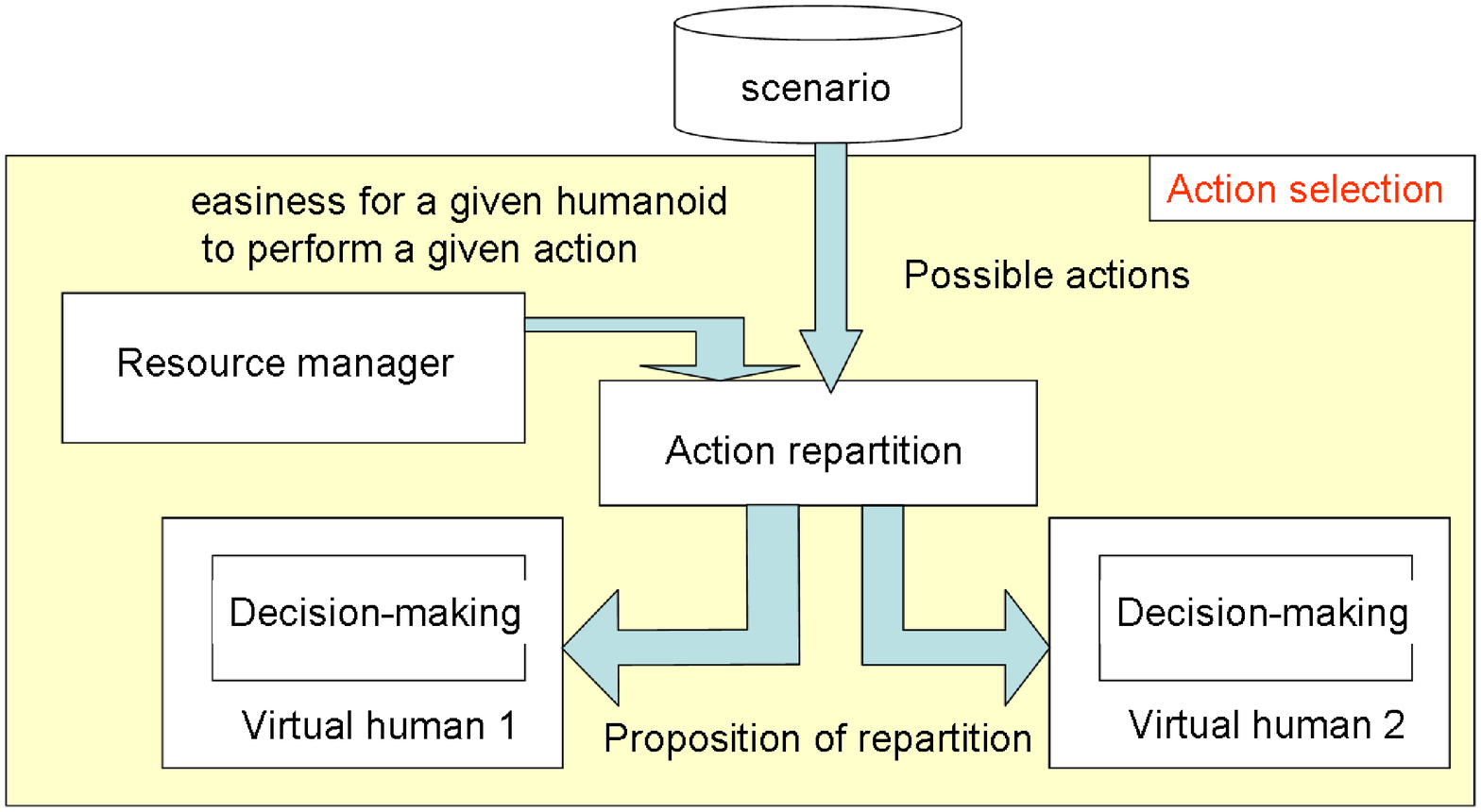}
\caption{The mechanism of action selection} 
\label{fig:selec}
\end{center}
\end{figure}

The module of action repartition makes a global repartition of the actions among the humanoids, respecting the scenario requirements. But it is only a propositional repartition and it is up to the virtual human to make an individual choice thanks to its own decisional module. These two modules form 
the whole process of "action selection", illustrated on \figurename \ref{fig:selec}, that we describe now.
\subsubsection{Action repartition}
\label{action repartition}
Several roles can be associated with one action, but this leads to a problem: how to evaluate who is the best candidate for this action. This evaluation will be done by this module of action repartition. This module aims at sorting the candidates for each possible action giving them a score according to various criteria such as the role priority, the proximity with the object to interact with, the easiness for the humanoid to do the action, etc. This last criterion will be determined thanks to a resource manager which has planification functionalities and which is able to predict blocking situations.
The criteria to take into account can be various but are easy to parametrize: we must attribute to each one a weight, possible values and a coefficient for each value. A positive coefficient indicates that this value tends to favour the choice of the associated humanoid.
The goal of this module is dual: the first one is to help a virtual human in his decisional process because, for each action allowed in the scenario, this module gives the best candidate from the scenario point of view. 
The second one is to propose a pedagogical feedback to the real user about the best action to choose.
To illustrate these goals, let us take an example: a virtual human that can do two actions in the scenario. The first, action 1, is easy for him because he is near the object to interact with, but a lot of people can do it. The second, action 2, is less easy, but in theory a second person could do it. Let us imagine that this second person can not do it, because he is busy for example. Without this module the virtual human would not have enough data to choose between these two actions, but with this module, he will see that for action 2 he is the only one who can do it, so he will deduce that it would be better from a global point of view to choose this one. 
As a result, this module makes propositions of repartition that is coherent with the scenario in order to help the humanoid to take a decision about the next action to do. 

\subsubsection{Decision-making} 
\label{decision-making}
The last step of the mechanism of action selection is the decisional module of a virtual human. Considering the results of the action repartition module, and the pedagogical profile of the virtual human, this module chooses the next action to do. The first step is to collect the actions provided by other modules. The interaction engine provides the actions that are possible. The action repartition module provides the actions allowed in the scenario, with information about the best actions to do. The pedagogy engine can provide specific demands, for pedagogical reasons. The second step is to tag these actions in order to identify specific ones like important actions (collaborative ones, urgent ones), hindering actions, etc. The final step is to respect the pedagogical profile of the virtual human to select the next action to do. The pedagogical profile is a set of propensities (for example to respect the procedure, to make mistakes, etc), it can correspond to the pedagogical roles proposed by Buche et al. \cite{Buche2003}.

\subsubsection{Synthesis}
In short, allowing several roles for one action leads to some interesting properties: flexibility, implicit collaboration, perfecting of scenarios. But in return, the global process of action selection must be divided into two parts: a global repartition which is optimized from the scenario viewpoint, and an individual choice made by each humanoid.

\section{Industrial application and validation: GVT project}
Our platform has been developed within the GVT project (Giat Virtual Training).
This project is a Research/Industry collaboration, with three partners: IRISA and CERV laboratories, and NEXTER-Group (previously Giat Industries). This last partner is an important French company, specialized in military equipments such as the Leclerc tank.
The main current application of our platform (called GVT) is virtual training on NEXTER's maintenance procedures.
GVT could be used, not only for maintenance procedures, but also for training on various kinds of procedures, such as diagnosis procedures, procedures of vehicle starting, etc. And nowadays, collaborative procedures are also possible. 
We are currently integrating our propositional models to support collaboration in GVT in order to validate them. 
As GVT leans on OpenMASK platform, we took advantage of its distribution possibility to distribute GVT. Now several distant visualizations can be created, allowing different users to connect on the same scene. For example the trainer can freely navigate in the scene while a trainee is training on a procedure.
The integration of humanoids is under progress: we can move avatars in the environment, and a first collaborative procedure with a virtual human has been created (\figurename \ref{fig:humano}). This procedure is the taking down of the auxiliary winch on a vehicle. Two roles are needed: an operator who realizes the main part of the procedure and an assistant who helps him to manoeuvre the winch. The trainee plays the role of the operator and a basic virtual human plays the role of the assistant. The virtual human have specific actions to do like pulling the cable (\figurename \ref{fig:humano2}). This action is automatically triggered when the trainee realizes the correct action that precedes in the procedure.

\section{Conclusion}
In our propositional models, we allow the association of several roles with one scenario action, and let the repartition module make propositions about the best candidates for each action. The repartition of the tasks is not fixed any more, and 
we believe that this will extend the possibilities offered by a CVE for training: flexibility of the scenarios, emergence of implicit collaboration between virtual humans and real users, possibility of perfecting procedures by letting virtual humans realize the procedure and by seeing the actions repartition.
Furthermore, we will be able to parametrize the behavior of our virtual humans thanks to their pedagogical profiles. Thus it will be possible to create virtual humans that simply assist the trainee performing the actions he has to, but also to create virtual humans who can make mistakes or who want to hinder the trainee. These possibilities will increase the pedagogical opportunities.

\subsubsection{\ackname}
The authors would like to thank all the members of the GVT project as well as Fabien Lotte, St\'ephane Donikian, St\'ephane Le Foll and Gr\'egoire Jacob for their helpful comments on this paper.

\bibliographystyle{unsrt}
\bibliography{biblio_edutainment.bib}

\end{document}